\documentclass[conference]{IEEEtran}
\IEEEoverridecommandlockouts
\usepackage{cite}
\usepackage{amsmath,amssymb,amsfonts}
\usepackage{algorithmic}
\usepackage{graphicx}
\usepackage{textcomp}
\usepackage{xcolor}
\usepackage{hyperref}

\def\BibTeX{{\rm B\kern-.05em{\sc i\kern-.025em b}\kern-.08em
    T\kern-.1667em\lower.7ex\hbox{E}\kern-.125emX}}
\begin{document}

\title{On the Study of Partial Qubit Hamiltonian for Efficient Molecular Simulation Using Variational Quantum Eigensolvers\\
}

\author{\IEEEauthorblockN{Harshdeep Singh}
\IEEEauthorblockA{\textit{CCDS} \\
\textit{IIT Kharagpur}\\
harshdeeps@kgpian.iitkgp.ac.in}
\and
\IEEEauthorblockN{Sabyashachi Mishra}
\IEEEauthorblockA{\textit{Department of Chemistry} \\
\textit{IIT Kharagpur}\\
mishra@chem.iitkgp.ac.in}

\and
\IEEEauthorblockN{Sonjoy Majumder}
\IEEEauthorblockA{\textit{Department of Physics} \\
\textit{IIT Kharagpur}\\
sonjoym@phy.iitkgp.ac.in}
}

\maketitle

\begin{abstract}
Quantum computing is being extensively used in quantum chemistry, especially in simulating simple molecules and evaluating properties like the ground state energy, dipole moment, etc. The transformation of a molecular Hamiltonian from the fermionic space to the qubit space provides us with a series of Pauli strings and the energy calculation involves the evaluation of the expectation values of all these individual strings. This introduces a major bottleneck for applications of VQEs in quantum chemistry. 
Unlike the fermionic Hamiltonian, the terms in a qubit Hamiltonian are additive and the present paper exploits this property to describe a new approach for extracting information from the partial qubit Hamiltonian of simple molecules to design more efficient variational quantum eigensolvers. In the partial (qubit) Hamiltonian approach (PHA), the qubit Hamiltonian is studied term-by-term to understand their relative contributions to the overall energy and a partial Hamiltonian is constructed with fewer Pauli strings that can resolve the entire Hamiltonian. With PHA, we can simulate molecules at a much lower computational cost with a truncated Hamiltonian. Additionally, the outcomes of the measurements with PHA quench the error due to noise introduced by the quantum circuits. We have also demonstrated the application of PHA  as an initialization technique, where the simple partial Hamiltonian can be used to find a suitable initial state for a more complex system. The results of this study have the potential to demonstrate the potential advancement in the field of quantum computing and its implementation in quantum chemistry.
\end{abstract}


\section{Introduction}
Variational Quantum Algorithms (VQAs)~\cite{b1} find their applications in a variety of different areas including quantum chemistry~\cite{b3, b4}, where these algorithms are used to simulate simple molecules and find some of their defining properties. In quantum chemistry, molecular problems are usually framed in a multi-dimensional fermionic space. For a molecular system with $N$ electrons and $M$ nuclei (with nuclear charge $Z$ and nuclear mass $M$), the Hamiltonian (in atomic units) can be written as,
\begin{eqnarray}
H&=& -\sum_{A=1}^{M} \frac{\nabla_A^2 }{2M_A} -\sum_{i=1}^{N} \left( \frac{\nabla_i^2}{2}  + \sum_{A=1}^{M} \frac{Z_A}{r_{iA}} \right)\nonumber\\
&&+ \sum_{j>i}^N\frac{1}{r_{ij}} + \sum_{B>A}^M\frac{Z_A Z_B}{R_{AB}},
\label{eq:2}
\end{eqnarray}
where the first two terms express the kinetic energies of the nuclei and the electrons, respectively. The other terms describe the potential energy due to electron-nuclei, electron-electron, and nucleus-nucleus interactions respectively. This expression can be further simplified by taking into account the fact that nuclei are nearly stationary compared to electrons, an approximation known as the Born-Oppenheimer approximation. Under the Born-Oppenheimer approximation, the Hamiltonian can be simplified to,
\begin{equation}
H=-\sum_{i=1}^{N} \left( \frac{\nabla_i^2}{2}  + \sum_{A=1}^{M} \frac{Z_A}{r_{iA}} \right) + \sum_{j>i}\frac{1}{r_{ij}}.
\label{eq:3}
\end{equation}
To solve a quantum chemistry problem on a quantum computer, the very first step is to reformulate the molecular Hamiltonian operator, which is defined in a fermionic space, in terms of the $2^N-$dimensional qubit space. While there are standard rules for this transformation, this conversion is easily understood when the quantum chemistry problem is expressed in the Fock space representation, also known as the second quantization notation, where the wave function can be written in terms of occupation numbers,
\begin{equation}
    \mid  k\rangle = \mid k_1,k_2, \cdots ,k_N\rangle,\quad k_p= \{0, 1\}.
    \label{eq:4}
\end{equation}
Here, $k_p = 0$ and $1$ signify the $p^{th}$ (spin)-orbital as unoccupied and occupied, respectively. In this representation of the wavefunction, the one-to-one correspondence between the fermionic and the qubit space is straightforward, i.e.,
\begin{equation}
    \mid k\rangle = \mid k_1,k_2, \cdots ,k_N\rangle \rightarrow \mid q\rangle = \mid q_1,q_2, \cdots ,q_N\rangle
    \label{eq:5}
\end{equation}
where, each orbital and its occupancy ($k_p= \{0,1\}$) is represented by the state of a qubit $q_p= \{\uparrow , \downarrow\}$.
The operators in the Fock-space representation are expressed in terms of the creation and annihilation operators~\cite{b5}, ($a^\dagger_p$ and $a_p$, respectively), defined by,
\begin{eqnarray}
    a^\dagger_{p} \mid k\rangle = (1-\delta_{k_p,1}) \Gamma^k_p \mid k_1,k_2, . ,1_p, \cdots, k_n\rangle \label{eq:6} \\
    a_p \mid  k\rangle = \delta_{k_p,1} \Gamma^k_p \mid k_1,k_2, . ,0_p,\cdots ,k_n\rangle. 
    \label{eq:7}
\end{eqnarray}
\noindent where
\begin{equation}
  \Gamma^k_p = (-1)^{ \sum_{m<p}k_m}.
\end{equation}
\noindent The molecular Hamiltonian in the second quantization notation can be represented as,
\begin{equation}
    H = \sum_{p,q} h_{pq} a_p^\dagger a_q + \frac{1}{2}\sum_{p,q,r,s}h_{pqrs}a_p^\dagger a_q^\dagger a_s a_r 
    \label{eq:8}
\end{equation}
where, $h_{pq}$ and $h_{pqrs}$ are the equivalent of the one and two-electron integrals. The one-electron integrals are obtained from the kinetic energy and electron-nuclei interactions while the two-electron integrals are obtained from the electron-electron interactions, defined with the help of some basis functions $\{X(\Vec{x})\}$ as,
\begin{equation}
    \displaystyle h_{pq} = \int d \vec{x} X_p^*(\vec{x})\left(-\frac{\nabla^2}{2}-\sum_{A}\frac{Z_{A}}{r_{A \vec x}}\right)X_q(\vec{x})
    \label{eq:9}
\end{equation}
and 
\begin{equation}
    h_{pqrs} = \iint d \vec{x_1}d \vec{x_2}\frac {X_p^*(\vec{x_1})X_q^*(\vec{x_1})X_r(\vec{x_2})X_s(\vec{x_2})}{r_{12}}.
\end{equation}

\noindent In the Fock space representation, the operators can be transformed to the qubit-space by various transformation schemes, such as the Jordan-Wigner~\cite{b6}, Parity~\cite{b8}, and Brayvi-Kitaev~\cite{b7} schemes. In Jordan-Wigner representation, the transformation rules are,
\begin{eqnarray}
    a_i^\dagger  &=& \frac{1}{2} (X_i - iY_i) \otimes_{j<i} Z_j \label{eq:tr_rule1}\\
    a_i &=& \frac{1}{2} (X_i + iY_i) \otimes_{j<i} Z_j \label{eq:tr_rule2}.
\end{eqnarray}
\noindent where, $a_i$, and $a_i^\dagger$ are the ladder operators defined in the fermionic space, and $X, Y, Z$ are the Pauli operators defined in the qubit space. With these transformation rules, the fermionic Hamiltonian (Equation \ref{eq:3} or \ref{eq:8}) can be transformed into an executable operator in the qubit space. The number of qubits and the Pauli strings in the qubit Hamiltonian would depend on the molecular system considered. For example, in the case of the H$_2$ molecule, the qubit Hamiltonian can be written as~\cite{b7},
\begin{equation}
    H_{\rm qubit}^{\rm H_2} = (-0.805) IIII + \cdots + (0.121) ZZII.
\end{equation}
This Hamiltonian is in a sum of products form, where each of its terms is an operator itself, and one of the objectives of this paper is to observe and understand the contribution and importance of each of these individual Pauli strings. The important question to answer would be if we can extract any useful information by considering the partial Hamiltonian, and if we can diagonalize the whole Hamiltonian by analyzing just some of its terms, making the overall computation smaller.\\

The PHA method can be summarized as follows: Consider a qubit Hamiltonian, consisting of n terms:
\begin{equation}
    H_{\rm full} = H_1 + H_2  + \cdots + H_n
\end{equation}

The variational quantum eigensolver finds a wavefunction that minimizes the energy,
\begin{equation}
   H_{\rm full}\mid \Psi_{\rm full}\rangle =E_{\rm full}\mid \Psi_{\rm full}\rangle 
\end{equation}

Now, with the PHA method, only a certain number of terms of the full Hamiltonian are considered, assuming that since the Hamiltonian is in sum-of-products format, some of the Pauli strings that don't contribute significantly to the final result can be ignored.
\begin{equation}
   H_{\rm partial} = H_1 + H_2  + \cdots  + H_m          (m<n) 
\end{equation}

and, 
\begin{equation}
   H_{\rm partial} \mid \Psi_{\rm partial}\rangle =E' \mid \Psi_{\rm partial}\rangle
\end{equation}

PHA method then examines whether the wavefunction that minimizes the partial Hamiltonian can also minimize the exact Hamiltonian of the system, or at least provide a result within chemical accuracy,
\begin{equation}
    \langle \Psi_{\rm full} \mid H_{\rm full} \mid \Psi_{\rm full} \rangle \stackrel{?}{=} \langle  \Psi_{\rm partial} \mid H_{\rm full} \mid \Psi_{\rm partial} \rangle
\end{equation}

\section{Potential Bottleneck for Quantum Computing in Chemistry Applications}
\noindent One of the potential impediments to the applicability of variational quantum algorithms in quantum chemistry is the sheer size of a qubit Hamiltonian of larger systems. This can be explained as follows, consider a qubit Hamiltonian of the form,
\begin{equation}
    H = H_1 + H_2  + \cdots + H_n
\end{equation}
Now, the energy measurement of this Hamiltonian can be achieved as,
\begin{equation}
  E = \langle \Psi \mid H\mid \Psi\rangle = \langle \Psi \mid [H_1 + H_2  + \cdots + H_n] \mid \Psi\rangle 
\end{equation}
\begin{equation}
   E = \langle \Psi \mid H_1\mid \Psi\rangle + \langle \Psi \mid H_2 \mid \Psi\rangle + \cdots + \langle \Psi \mid H_n\mid \Psi\rangle   
\end{equation}
Therefore, the energy measurement requires the expectation values of each of these Pauli strings, whose number increases rapidly with the enhancement in the number of qubits, which becomes a problem for larger molecular systems. The number of Pauli strings in a qubit Hamiltonian scale with the number of qubits in the system as $ O(n^4) $. This is where the PHA can be extremely useful as it might be possible to find a much smaller set of Pauli strings required to diagonalize the Hamiltonian.

%\begin{figure}[htbp]
%\includegraphics[width=0.5\textwidth]{Number_of_terms.png}% Here is how to import EPS art
%\caption{\label{fig:fig1}The scaling of the number of the Pauli Strings in the Hamiltonian of molecular systems with an increase in the number of Qubits.}
%\end{figure}

\section{Previous Works on Cost Reduction}
There have been a bunch of theoretical~\cite{b11, b12, b13} and experimental~\cite{b14, b15} developments in the applications of quantum computing research related to quantum chemistry over the years, and work has been done to reduce the cost of circuit execution~\cite{b16}. One such closely related work involves term-truncation by exploiting spatial locality along with the Bravyi-Kitaev transformation~\cite{b17}, which tries to reduce the overall cost by reducing the number of measurements and state preparations that are not expected to contribute at the desired precision to the final result. However, the existing method has been studied extensively for Quantum Phase Estimation (QPE) but has not been extended to VQEs. This method aims to reduce the cost by employing different basis functions and exploring the contributions of the one-and two-electron integrals. With PHA, we try to explore ways to reduce the simulation cost post-fermion-to-qubit transformation and present a much simpler method to efficiently simulate the quantum system. As it will be highlighted later, PHA also allows to add corrections to the final results and can be extended to be used as an initialization method for the VQE, making it a bit more versatile.

\section{Criteria for Constructing the Partial Hamiltonian}

\noindent The major testing and experimentation have been done on the Hydrogen molecule, in its full form, defined with four qubits and 15 terms, as highlighted in TABLE~\ref{tab1}. The small size of the Hydrogen molecule allows for multiple simulation runs and makes it easy for hit-and-trial methods to define a general scheme for constructing the partial Hamiltonian. All molecular energies are reported in the atomic units without including the nuclear repulsion energy, which is a constant under the Born-Oppenheimer approximation. 

\begin{table}[htbp]
\caption{Hydorgen Molecule Qubit Hamiltonian}
\begin{center}
\begin{tabular}{|c|c|c|c|c|c|}
%\hline
%copy& More table copy$^{\mathrm{a}}$& &  \\
\hline

Term & Coefficient  & Gates & Term & Coefficient  & Gates\\
\#$^{\mathrm{a}}$ & & & \#$^{\mathrm{a}}$ & & \\
\hline

1    &  $-$0.81054  & IIII  & 9     & $-$ 0.04523 & IXZX\\
2    & $+$0.17218   & IIIZ  &  10   & $-$ 0.04523& ZXZX\\
3    & $-$0.22575   & IIZZ  &  11   & $+$ 0.04523 & IXIX\\
4    & $+$ 0.17218   & IZZI  &  12   & $+$ 0.16614 & ZZIZ\\
5    & $-$ 0.22575   & ZZII  &  13   & $+$ 0.16614 & IZIZ\\
6    & $+$ 0.12091   & IIZI  &   14  & $+$ 0.17464 & ZZZZ\\
7    & $+$ 0.16892   & IZZZ  &  15   & $+$ 0.12091 & ZIZI\\
8    & $+$ 0.04523  & ZXIX  & & & \\
%\botrule
\hline
\multicolumn{6}{l}{$^{\mathrm{a}}$The sequence follows the Qiskit ordering.}
\end{tabular}
\label{tab1}
\end{center}
\end{table}

\begin{figure}[htbp]
\includegraphics[width=0.5\textwidth]{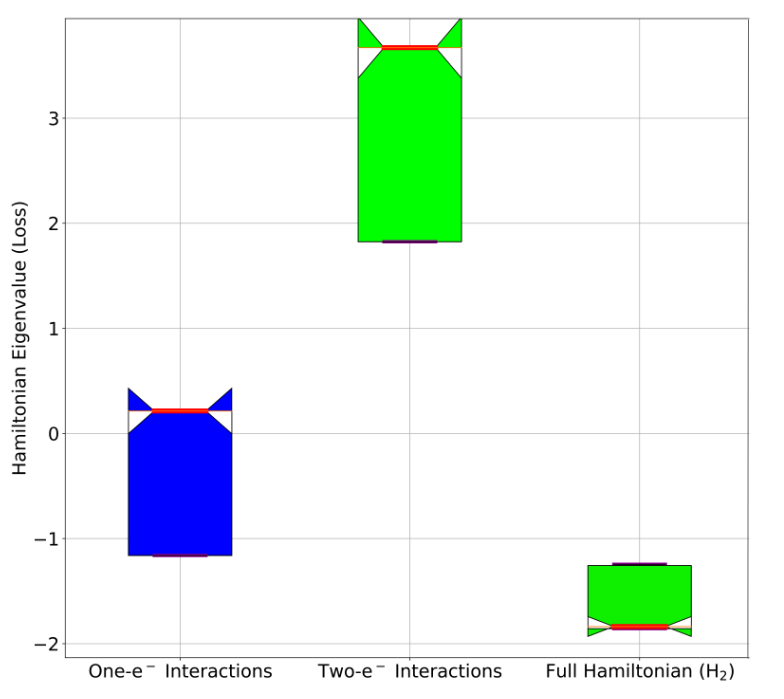}% Here is how to import EPS art
\caption{\label{fig:fig2}The results of 100 VQE simulations for Hydrogen Molecule (in Hartree Units) when only One and Two Electron Interactions are Considered Individually Against the Full Hamiltonian.}
\end{figure}

\noindent The qubit Hamiltonian is constructed with two different parts- the one-electron interactions (kinetic energy of individual electrons and the electron-nuclei interactions) and the two-electron interactions (electron-electron interactions). Figure~\ref{fig:fig2} shows how VQE would perform if only one, or, two-electron contributions are considered, and based on that, it is clear that ignoring one set of interactions altogether would not work and some combination of both interactions is required.

\begin{figure}[htbp]
\includegraphics[width=0.5\textwidth]{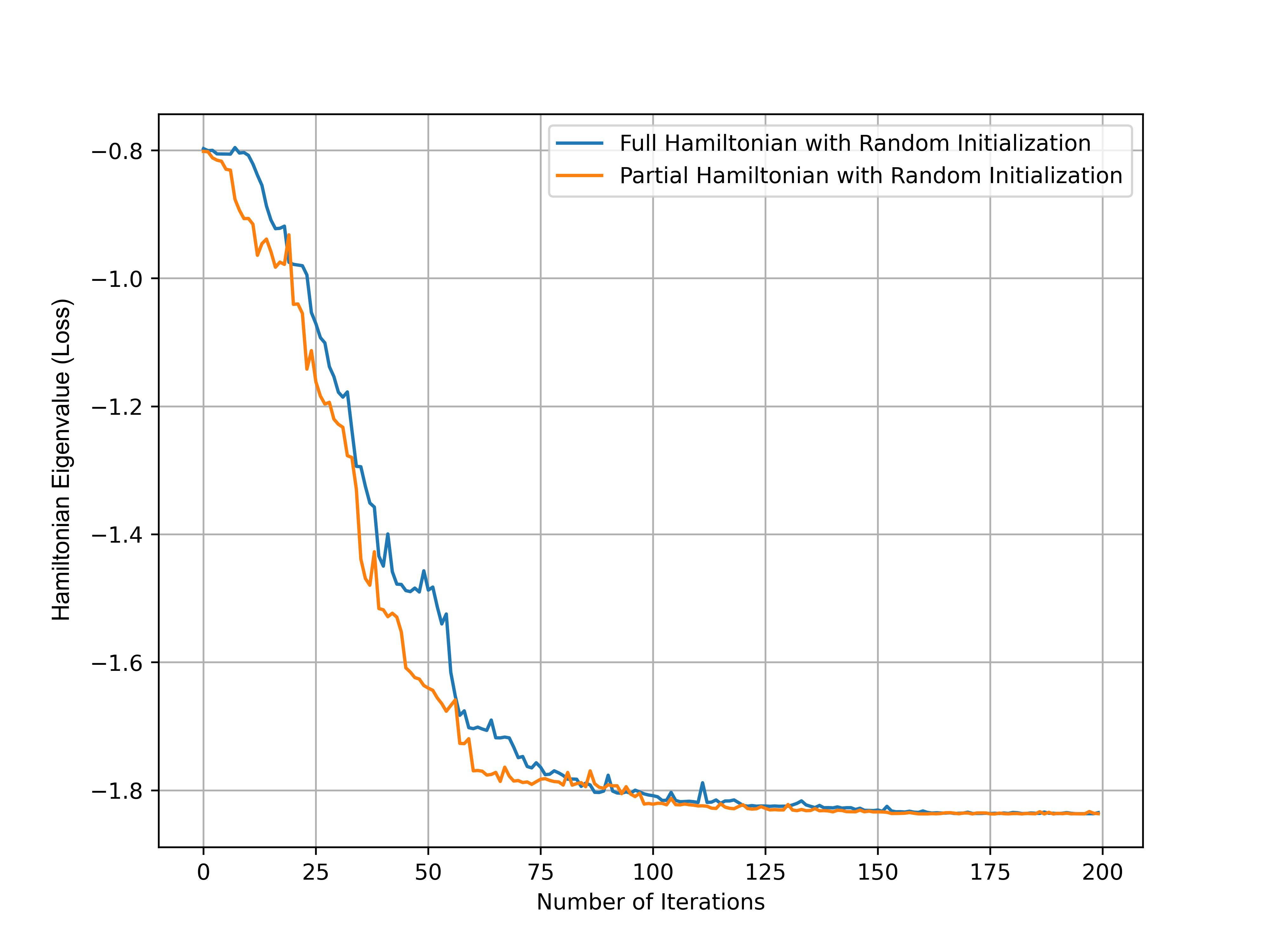}% Here is how to import EPS art
\caption{\label{fig:fig3}The result of Hydrogen Molecule simulation for the full and partial Hamiltonian (in Hartree Units). Both simulations were carried out with the same randomly-initialized initial state.}
\end{figure}

\begin{figure}[htbp]
\includegraphics[width=0.5\textwidth]{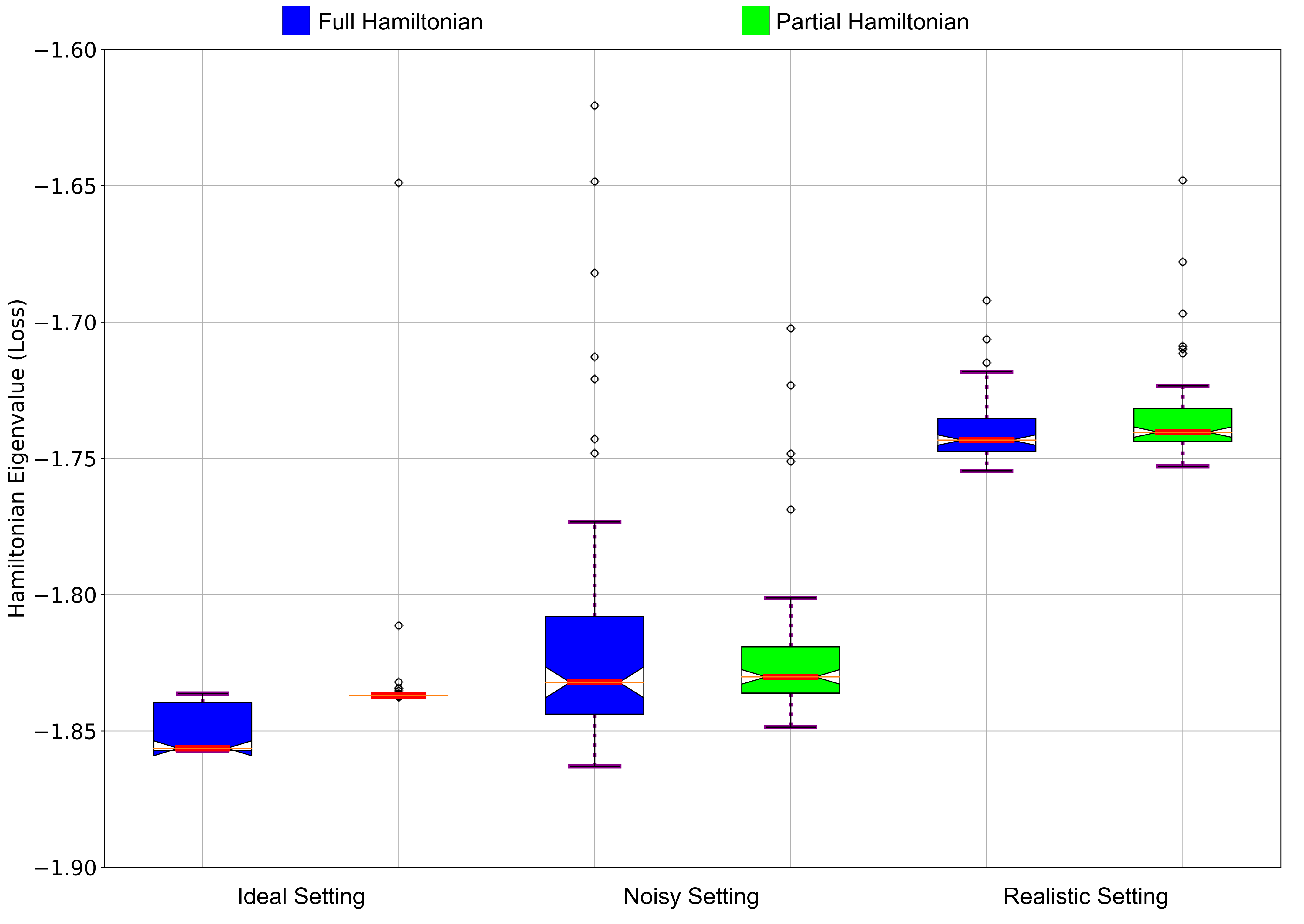}% Here is how to import EPS art
\caption{\label{fig:fig3a}The results of 100 VQE simulations for Hydrogen Molecule for full and partial Hamiltonians (in Hartree Units) with three different settings: Ideal Quantum Circuit Simulator (Ideal Setting), Noisy Quantum Circuit Simulator (Noisy Setting), Noisy Quantum Circuit Simulator with Noise Embedded from IBM Cairo Device (Realistic Setting).}
\end{figure}

\noindent The partial Hamiltonian can be constructed in a manner that requires the minimum number of circuits to be executed, and one way of doing that is just to consider the terms with either 'I', or 'Z' gates, as a single circuit would then be sufficient to evaluate the result. The result in Fig.~\ref{fig:fig3} highlights the convergence performance of the VQE for the full and the partial Hamiltonian of the Hydrogen molecule, $E_{\rm full}$, and $E_{\rm partial}$, where
\begin{equation}
    E_{\rm full} = \langle \Psi_{\rm full} \mid H_{\rm full} \mid \Psi_{\rm full} \rangle
\end{equation}
and, 
\begin{equation}
    E_{\rm partial} = \langle \Psi_{\rm partial} \mid H_{\rm full} \mid \Psi_{\rm partial} \rangle
\end{equation}
where, $|\Psi_{\rm full} \rangle$, and $|\Psi_{\rm partial} \rangle$ are the states achieved by running VQE on the full and partial Hamiltonian, and is the state achieved by running VQE on the partial Hamiltonian. While this result ensured the convergence of the algorithm, the same simulation was carried out a number of times (100) with different settings to confirm that the results are robustly valid throughout the following cases, an ideal quantum circuit simulator, a noisy quantum circuit simulator, and then lastly, the noisy quantum circuit simulator with external noise embedded from IBM Cairo Device, as highlighted in Fig.~\ref{fig:fig3a}. In ideal simulation settings, CG optimizer was employed while in case of noisy settings, SPSA was used, based on our previous work~\cite{bh}. The results in Fig.~\ref{fig:fig3a} are presented in a notched box-plot~\cite{bhh}, which gives a clear idea of the spread of the VQE energies over 100 simulation runs, highlighting the median energy, the effect of noise, and local minima on the performance of the algorithm. It can be observed that the average energy across multiple simulations with both the full and partial Hamiltonian is in excellent agreement and this observation was replicated in all sets of simulations. Hence, with the data so far, it can be assumed that the VQE does not require resolving the entire Hamiltonian of a molecule, rather than, just considering a much simpler version of its Hamiltonian (partial Hamiltonian) with a much lower quantum and classical computational cost. 

As a result, the requirements for building the partial Hamiltonian of a molecule can be stated as follows:
\begin{enumerate}
    \item It is necessary to take into account the term in the full Hamiltonian with the largest co-efficient, typically this is the identity term, which is constant.
    \item All terms with either the 'I', 'Z', or both gates must be taken into account.
    \item Depending on the coefficient value, more terms may be introduced if convergence is not obtained. The terms with higher coefficients must be taken into account first.
\end{enumerate}

\begin{figure*}[htbp]

\centerline{\includegraphics[width=1.0\textwidth]{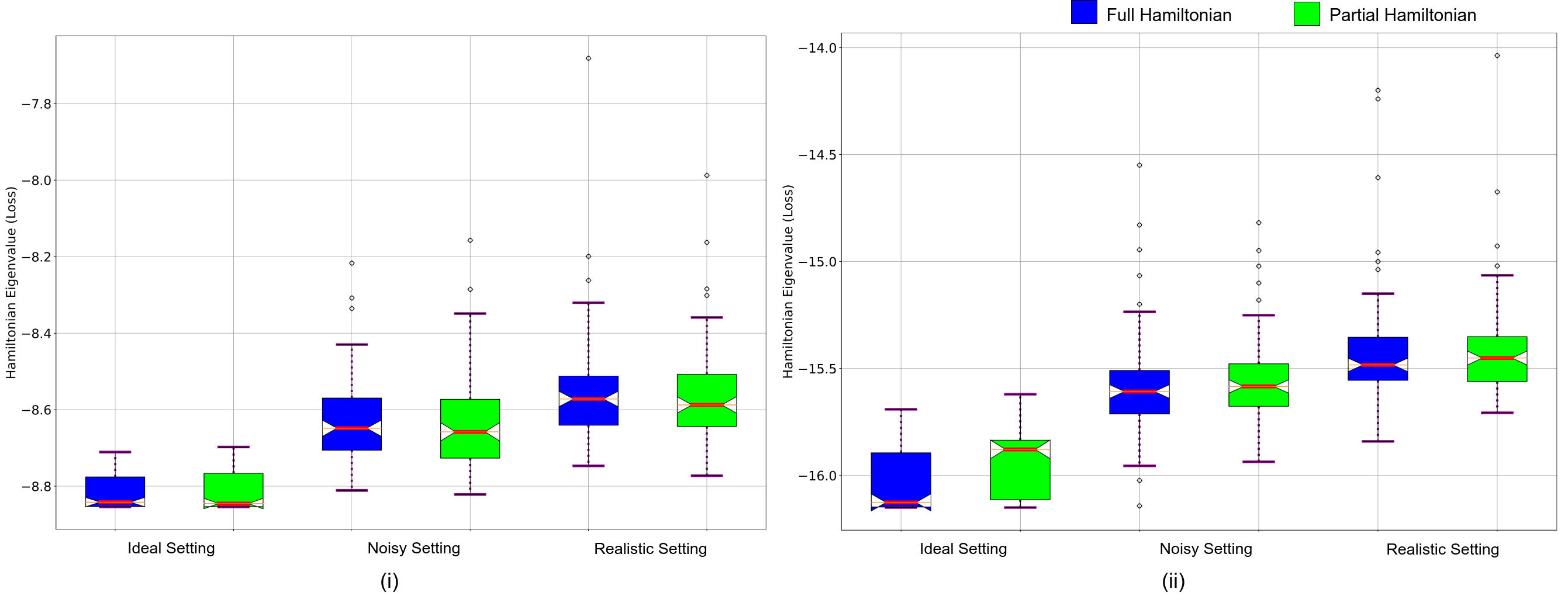}}
\caption{The results of 100 VQE simulations for (i) Lithium Hydride, and (ii) Berrylium Hydride for full and partial Hamiltonians (in Hartree Units) with three different settings: Ideal Quantum Circuit Simulator (Ideal Setting), Noisy Quantum Circuit Simulator (Noisy Setting), Noisy Quantum Circuit Simulator with Noise Embedded from IBM Cairo Device (Realistic Setting).}
\label{fig:fig4}
\end{figure*}

Based on these criteria, the simulation results for Lithium Hydride and Berrylium Hydride are presented in Fig.~\ref{fig:fig4} and it can be seen that the PHA returns excellent agreements with the full Hamiltonian approach. The average energies (in Hartree units) over 100 simulation runs for different molecules, with system size ranging from 4 qubits (in Hydrogen molecule) to 10 qubits (in Water molecule) are presented in TABLE~\ref{tab2}.

\begin{table}[htbp]
\caption{Simulation Results for Different Molecules}
\begin{center}
\begin{tabular}{|c|c|c|c|c|}
%\hline
%copy& More table copy$^{\mathrm{a}}$& &  \\
\hline

Molecule & Method & \# Qubits & Full  & PHA$^{\mathrm{a}}$\\
 &  & & Hamiltonian$^{\mathrm{a}}$ & \\
\hline

H$_2$ & Ideal & 4    & $-$1.82781 & $-$1.79306 \\
      & Noisy  &   &  $-$1.76914  & $-$1.73434\\
      \hline
LiH   & Ideal & 6    & $-$8.79805 & $-$8.81813 \\
      & Noisy  &  &   $-$8.50216& $-$8.46643 \\
        \hline
BeH$_2$  & Ideal  & 8     & $-$15.14592 & $-$15.12351\\
      & Noisy  &  &  $-$14.79626 & $-$14.70781\\
        \hline
H$_2$O   & Ideal & 10    & $-$23.51381 & $-$23.65410\\
         & Noisy &    & $-$22.49302 & $-$22.39352\\

%\botrule
\hline
\multicolumn{5}{l}{$^{\mathrm{a}}$Averaged over 100 different simulation runs, in Hartree units.}
\end{tabular}
\label{tab2}
\end{center}
\end{table}

\subsection{The Effect of Noise in Simulation Results with Exact and Partial Hamiltonians}
\begin{figure*}[!htbp]
\centerline{\includegraphics[width=1.0\textwidth]{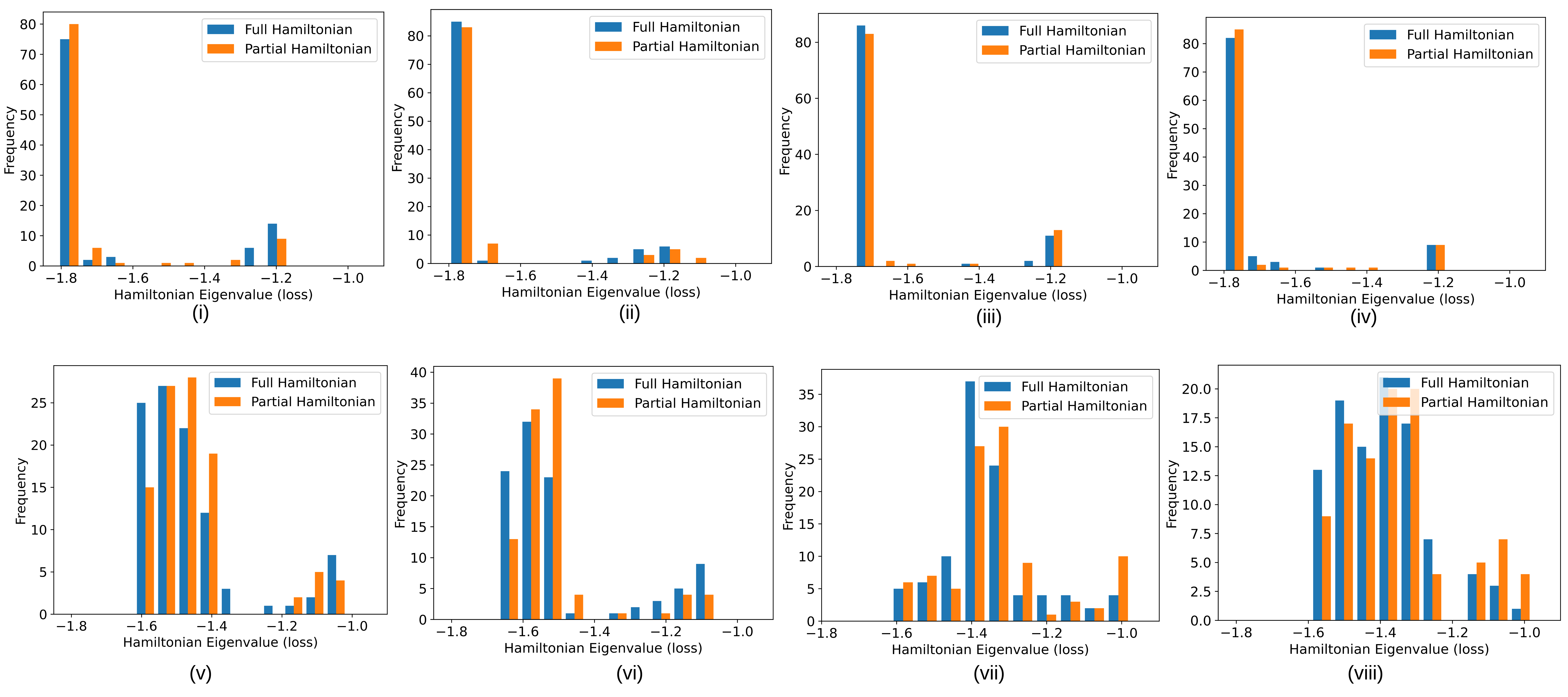}}
\caption{The frequency distribution of Hydrogen Molecule energy over 100 VQE simulations with full and partial Hamiltonians (in Hartree Units) with 
Noisy Simulator and the noise embedded from the following IBM devices: (i) Vigo, (ii) Bogota, (iii) Belem, (iv) Athens, (v) Quito, (vi) Perth, (vii) Nairobi, and (viii) Lima.}
\label{fig:fig5}
\end{figure*}

\noindent One of the major concerns in the application of the partial Hamiltonian method is what effect noise would play in the overall performance. While Fig.~\ref{fig:fig3a} does present results from a realistic quantum setting where the hydrogen molecule simulation was carried out in a noisy quantum circuit simulator with external noise embedded from the IBM Cairo device. However, for generalization, the same was then carried out with multiple noise models and the frequency distribution of the energy can be found in Fig.~\ref{fig:fig5}, where it is evident that the PHA performs quite well in all the different simulators and with different noise models used.

\section{Partial Hamiltonian as an Initialization Method for VQEs}
\begin{figure}[htbp]
\centerline{\includegraphics[width=0.5\textwidth]{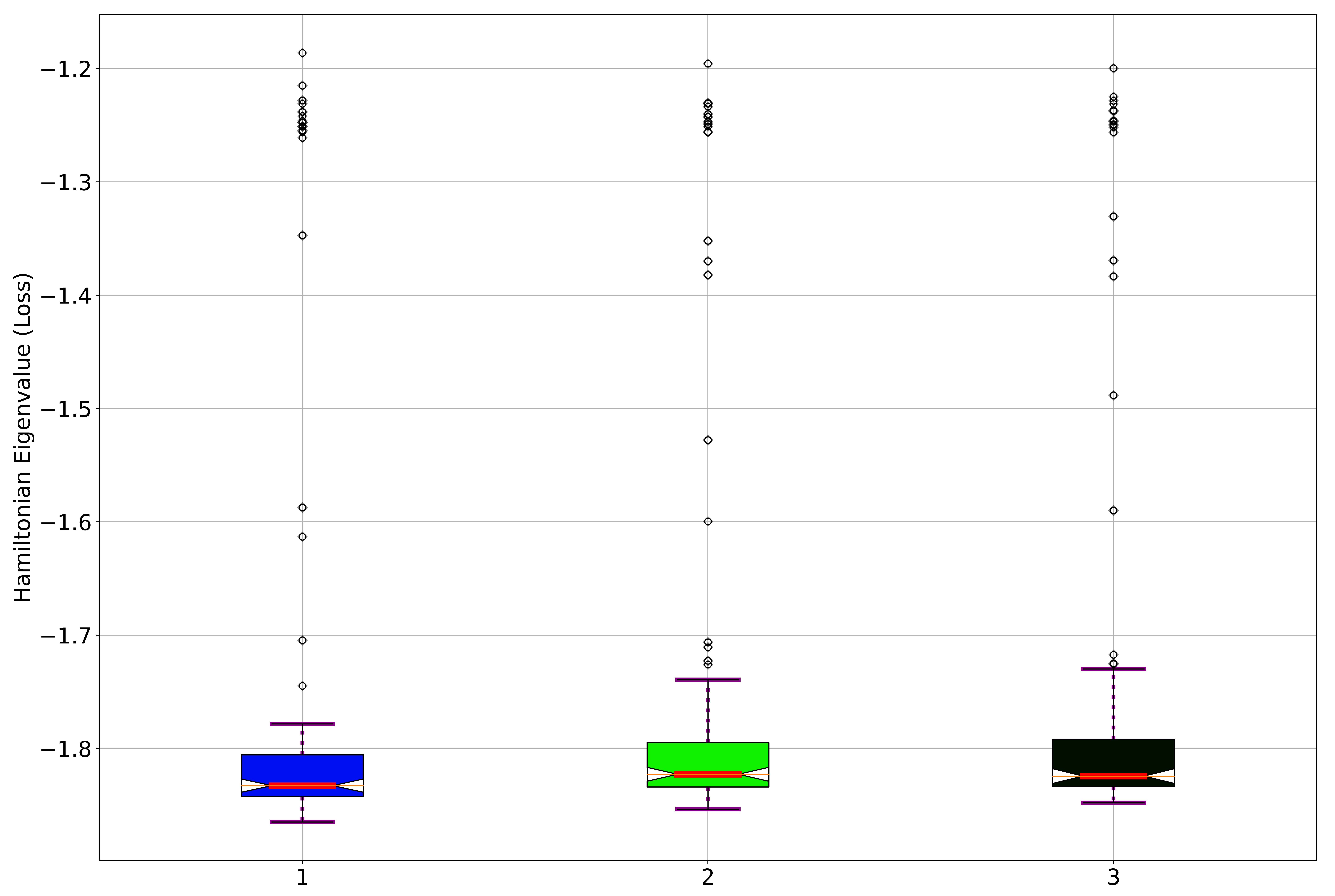}}
\caption{The results of a 100 VQE simulations for Hydrogen Molecule (in Hartree Units) with (1) Full Hamitonian, (2) Partial Hamiltonian, and (3) Full Hamiltonian with Partial Hamiltonian Initialization}
\label{fig:fig6}
\end{figure}

\noindent With the PHA method, there will always be concern over the exact result of the simulation since a majority of the terms of the molecule's Hamiltonian are not been considered and as the system size grows larger, the ignored Pauli strings might end up contributing more and more. One possible solution to this problem would be to use the PHA as an initialization method, that is, the partial Hamiltonian can be used to find a good initial state for the overall system. This has a huge potential upside as the bulk of the simulation can be carried out on a simpler Hamiltonian. An initial trial was done with the Hydrogen molecule, see FIG.~\ref{fig:fig6}, and while the difference is not apparent, it was found that the average energy found over 100 simulations with partial Hamiltonian initialization became almost the same as the full Hamiltonian. 

\section{Discussions and Potential Computational Advantage}

\noindent PHA can deliver potential computational advantage considering the fact that all the strings in the partial Hamiltonian containing either the `I', or, `Z' gate commute with each other. This implies that they can be grouped together and a single measurement would be sufficient to calculate the energy. This could be a massive gain, considering the fact that the number of Pauli strings in the qubit Hamiltonian scale as $O(n^4)$, and then the number of shots required for an accuracy of $\epsilon$ must be $O(n^4/\epsilon^2)$~\cite{b10}. The number of shots on employment of the partial Hamiltonian can be reduced to $O(1/\epsilon^2)$. The exact computational advantage of this method requires further study with bigger molecules and also when it is being used as an initialization technique.\par

Another potential advantage that the partial Hamiltonian would offer courtesy to the previous discussions is that since the number of Pauli strings is smaller, the number of circuit elements and therefore, the overall number of measurements can be significantly reduced. This can improve the performance of the VQEs against the noise introduced by the quantum circuits. This is also shown in the results so far, as the overall spread of the simulation results appears to be smaller for partial Hamiltonian than the full Hamiltonian and the number of outliers is also smaller, see FIG.~\ref{fig:fig3}, and FIG.~\ref{fig:fig4}.
%\section{Potential Computational Advantage}
\begin{figure}[htbp]
\centerline{\includegraphics[width=0.5\textwidth]{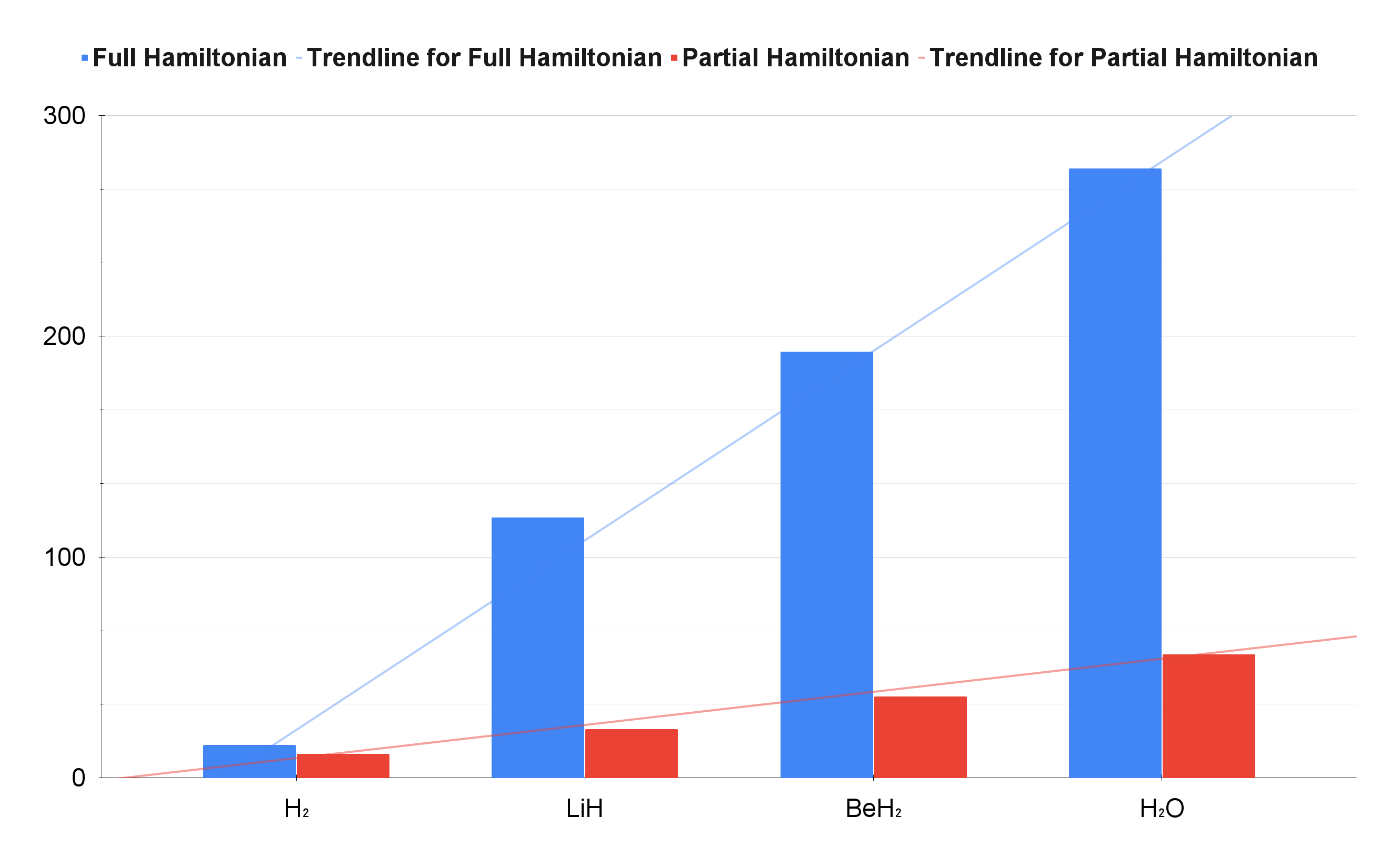}}
\caption{Number of Pauli Strings in the full and partial Hamiltonian of different molecules, and the trendline of the order with increasing system size.}
\label{fig:fig7}
\end{figure}

\section{Conclusion}
Data from FIG.~\ref{fig:fig3}, and FIG.~\ref{fig:fig4} strongly suggest that the PHA method provides an efficient way of molecular simulation, with a massive decrease in the length of Pauli strings that need to be measured for the energy calculation. FIG.~\ref{fig:fig7} shows the difference in the number of terms in the full and partial Hamiltonian of the molecules considered for simulation in this work, and it shows that one can get pretty reasonable results by considering a significantly smaller Hamiltonian. The performance of the method against different noise models was a particularly great result, see FIG.~\ref{fig:fig5}. Based on this and the discussions earlier, the partial Hamiltonian method appears to be a great counter to the bottleneck introduced by the large size of molecular systems.  

\section{Future Directions}
There is still plenty of directions in which one can move forward with the partial Hamiltonian analysis. For one, the performance of the method with larger and more complicated systems needs to be studied. Using this method as an initialization method can also yield significant speedups and remains an open area of research. This method has some resemblance to the feature selection method widely used in machine learning algorithms, as choosing which Pauli strings to keep in the partial Hamiltonian is similar to choosing which features to be selected in the training process, so that can be a potential crossover between machine learning and quantum chemistry applications. 
%A study of the cost landscape for partial Hamiltonian, similar to the work of~\cite{b18} also could help in constructing a more robust partial Hamiltonian, and therefore more efficient VQEs. 

\section*{Acknowledgment}

\noindent This work used the Supercomputing facility of IIT Kharagpur established under the National Supercomputing Mission (NSM), Government of India, and supported by the Centre for Development of Advanced Computing (CDAC), Pune. HS acknowledges the Ministry of Education, Govt. of India, for the Prime Minister's Research Fellowship (PMRF).

\end{document}